\documentclass[aps,pra,preprint]{revtex4}
\usepackage{amsmath}
\usepackage{amsfonts}

\begin{document}

\title{Entangled Brachistochrone: 
Minimum Time to reach Target Entangled State}
\author{Arun K.\ Pati, Biswajit Pradhan and Pankaj Agrawal}

\address{Institute of Physics, Bhubaneswar-751005, Orissa, India }

 \date{\today}

\begin{abstract}
We address the question: Given an arbitrary initial state and a general 
physical interaction what is the minimum time for reaching a target 
entangled state? We show that the minimum time is inversely proportional 
to the quantum mechanical uncertainty in the non-local Hamiltonian. We find 
that the presence of initial entanglement helps to minimize the waiting time.
Furthermore, we find that in a bi-local rotating frame the entangling 
capability is actually a geometric quantity.
We give an universal bound for the time average of entanglement rate for 
general quantum systems. 
The time average of entanglement rate does not depend on the 
particular Hamiltonian, rather on the fluctuation in the Hamiltonian. 
There can be infinite number of nonlocal Hamiltonians which may give same 
average entanglement rate. We also prove a composition law for minimum time 
when the system evolves under a composite Hamiltonian.
\end{abstract}



\maketitle

\section{Introduction}
Quantum entanglement plays a pivotal role in the emerging field of quantum 
information theory \cite{divin}. 
Creation, storage and processing of entangled states are
challenging experimental tasks in any quantum information processing devices
\cite{expt,expt1,expt2}.
In recent years there have been considerable progress in understanding 
various types of entangled states. However, in multiparticle situations, 
many questions are still unexplored. We know that if we start from an 
unentangled state of two (or more) particles and allow them to interact, 
then the state may evolve 
to an entangled state depending on the initial state and the type of 
interaction Hamiltonian. The Hamiltonian that is capable of creating 
entanglement is not sum of local Hamiltonians, rather it involves a 
nonlocal part. This is so, because sum of the local Hamiltonians will always 
give rise to tensor product of local 
unitary evolution operators and we know that the later cannot create any 
entanglement. In some situation, one would like to generate a 
particular entangled state for some specific quantum information 
processing task. For example, if one could create a maximally entangled state 
in a controlled way, then that can be used for quantum teleportation, dense 
coding, remote state preparation and so on.
However, what type of initial state and what kind of
non-local Hamiltonian can give the desired entangled state is far from clear.
Also, it is not the case that immediately after switching on the interaction 
Hamiltonian we will have the target entangled state.

There has been considerable interest in the study of dynamics of entanglement.
In particular, a pertinent question in this context is given a nonlocal 
Hamiltonian, how well one can create entangled states. This has been address by
D{\"u}r {\it et al} \cite{dur}. They have introduced entangling capability of 
Hamiltonian for two qubits and also generalized for two qudits.
It was found that it is better to start with initial entangled state and the 
best initial entanglement is independent of the physical process.
Also, it was shown that one can improve the capability if we allow 
fast local operations and ancillas (in some cases) \cite{dur}. 

In this paper we address the following question: Given an arbitrary 
product state and a general physical interaction what is the 
minimum time for creating a target entangled state? This is precisely the 
entangled brachistochrone problem. This question is not 
only of fundamental interest but also of practical importance. 
The answer will be relevant in many experimental context where we want a 
fixed entangled state after having the interaction on. Because, that will allow
us to generate the desired entangled state in a controlled way. This will also
answer the question what is the minimum time an experimentalist needs to wait 
to create the target entangled state. In the course 
of our investigation, we show that the entanglement capability of 
non-local Hamiltonains is a geometric quantity.
We prove a universal bound on the {\em time average of the entanglement rate}. 
Our result suggests that even though we cannot tell whether entangling rate and 
disentangling rates are same, but 
the maximum value of the time average of entanglement and disentanglement 
rates are the same.

We prove that the minimum time for creating a fixed entangled state not only 
depends on the initial state and final state, but also depends on the speed at 
which system evolves. The speed of the quantum evolution is governed by the 
fluctuation in the Hamiltonian $\Delta H$. In particular, we show that 
the minimum time required to reach a target entangled state 
depends inversely on the quantum mechanical uncertainty in 
the non-local Hamiltonian. We illustrate this for two qubits. 
Further, we will show that the time average of 
entanglement rate is 
upper bounded by $ 2 \log N \frac{ \Delta H}{\hbar S_0}$, where $N$ is the 
number of Schmidt coefficients in the entangled state, 
$\Delta H$ is the fluctuation in the 
nonlocal Hamiltonian and $S_0$ is the shortest geodesic connecting the 
initial and the final states. As an application of 
this bound we estimate what could be the ultimate entanglement rate for 
non-local Hamiltonians.

The present paper is organized as follows. In Sec. II, we briefly discuss 
the geometric uncertainty relation and the bound for the minimum time 
for entangled states. In Sec. III, we illustrate the bound for two qubits. 
We find that initial entanglement can help to reduce the waiting time 
in reaching a maximal entangled state. In Sec. IV, we show that entanglement
rate in a bilocal rotating frame is actually related to the quantum fluctuation 
in the nonlocal Hamiltonian (the speed of the quantum evolution). Also, we 
provide a universal bound for the time average of the entanglement rate.
In Sec. V, we discuss the question of minimum time when the system is driven
by composite Hamiltonian. Finally, we conclude our paper in Sec. VI.

\section{Geometric Uncertainty Relation and Minimum Time}

Here, we consider bipartite systems but some of the results will be valid for
more general situations. Whenever the result holds only for bipartite systems,
 we will say so. Consider an arbitrary initial product state $|\Psi(0) \rangle =
|\psi(0) \rangle \otimes |\phi(0) \rangle$ of a bipartite composite system. 
The nonlocal Hamiltonian for 
the system is given by $H = H_1 \otimes I + I \otimes H_2 + H_{\rm int}$. 
Under the action of the 
Hamiltonian the state evolves unitarily as $|\Psi(0) \rangle \rightarrow 
|\Psi(t) \rangle = \exp(-iH t /\hbar)|\Psi(0) \rangle$. Now, depending on the 
type of initial state and $H_{\rm int}$ the resulting state will be entangled.
Suppose, we evolve the system from $|\Psi(0) \rangle \rightarrow 
|\Psi(T) \rangle$ and 
 our target entangled state is $|\Psi_T \rangle = |\Psi(T) \rangle$.
Then, the question is what is the minimum time $T_{\rm min}$ 
for which we should evolve the combined system to get the desired state?

To answer this, we need some geometric ideas. Let $\{\Psi \}$ be a set of 
vectors in ${\cal H} =
{\cal H}_1 \otimes {\cal H}_2$. If these vectors are not normalized we can 
consider a set of vectors $\{ \Psi/|| \Psi|| \}$ of norm one in 
${\cal L}$. The set of 
rays of ${\cal H}$ is called the projective Hilbert space  
${\cal P}({\cal H}_1 \otimes {\cal H}_2)$.
If dim${\cal H}_1= N $ and dim${\cal H}_2= M$, then ${\cal H} \simeq 
{\bf C}^{N M}$. The projective Hilbert space is ${\cal P} = 
({\bf C}^{N M} - \{0\})/U(1)$ which is a complex manifold of dimension
$(N M -1)$. This can also be considered as a real manifold 
of dimension $2(N M -1)$. Any quantum state 
at a given instant of time can be represented as a point in ${\cal P}$ via
the projection map $\Pi: |\Psi\rangle \rightarrow |\Psi\rangle \langle \Psi|$. 
The evolution of the state vector can be represented by a curve
 $C: t \rightarrow |\Psi(t)\rangle$ in ${\cal H}$ whose projection 
$\Pi(C)$ lies in ${\cal P}$. Here, smooth mappings 
$C: [0, t] \rightarrow {\cal L}$ of an interval into a differentiable 
manifold are called smooth curves in the given manifold \cite{akp}. 

Now consider the unitary time evolution of a bipartite quantum system
$|\Psi(0) \rangle \rightarrow 
|\Psi(t) \rangle = \exp(-iH t /\hbar)|\Psi(0) \rangle$, where $H$ is the 
nonlocal Hamiltonian.
If the state at later time is entangled, then using the Schmidt decomposition 
theorem we can write the combined state as 
\begin{eqnarray}
|\Psi(t) \rangle = 
\sum_{n=1}^N \sqrt{\lambda_n(t)} |a_n(t)\rangle |b_n(t) \rangle, 
\end{eqnarray}
where $\lambda_n(t)$'s are the Schmidt coefficients with 
$\sum_n \lambda_n(t) =1$ for all time; $|a_n(t)\rangle$ and 
$ |b_n(t) \rangle$ are the orthonormal Schmidt basis. As the state 
evolves, it traces a path in the state space ${\cal P}$.
There is a natural metric on ${\cal P}$ which is the Fubini-Study metric, 
defined as 
\cite{aa,akp,akp1}
\begin{eqnarray}
dS^2 &=& 4 (1 - |\langle \Psi(t)|\Psi(t+dt) \rangle |^2) = 4 \frac{\Delta H^2}
{\hbar^2}dt^2,
\end{eqnarray}
where $\Delta H^2 = \langle \Psi(t)|H^2|\Psi(t) \rangle -  
\langle \Psi(t)|H|\Psi(t) \rangle^2$ is the quantum mechanical uncertainty in
the nonlocal Hamiltonian. For time-independent Hamiltonians $\Delta H$ is also 
time independent (because $H$ and $U(t)$ commutes). 
Using the notion of infinitesimal distance, 
one can define the rate or the speed $V$ at which the quantum system evolves in 
the projective Hilbert space. This is given by 
$V =  \frac{dS}{dt} = 2 \frac{\Delta H}{\hbar}$. We can write down an explicit
expression for the speed as (with the initial state as the product state) 
\begin{eqnarray}
V^2 = \frac{4}{\hbar}[ \Delta H_1^2 + \Delta H_2^2 + \Delta H_{\rm int}^2 + 
C(H_1, H_{\rm int}) + C( H_{\rm int}, H_1) + C(H_2, H_{\rm int}) + 
C(H_{\rm int}, H_2) ],
\end{eqnarray}
where  $\Delta H_1^2 = \langle \psi(0)|H_1^2|\psi(0) \rangle -  
\langle \psi(0)|H_1|\psi(0) \rangle^2$, $\Delta H_2^2 = 
\langle \phi(0)|H_2^2|\phi(0) \rangle -  
\langle \phi(0)|H_2|\phi(0) \rangle^2$, and $C(A, B)= 
\langle \Psi(0)|AB|\Psi(0) \rangle -  
\langle \Psi(0)|A|\Psi(0) \rangle \langle \Psi(0)|B|\Psi(0) \rangle$ being the 
correlation between $A$ and $ B$ (here $A$ and $B$ can be $H_1$, $H_2$ or 
$H_{\rm int}$).

We can also give an expression for the speed of the entangled state evolution 
using the Schmidt basis and coefficients.
Now, using (1) the Fubini-Study entangled metric can be expressed as
\begin{eqnarray}
dS^2 & = & \sum_n \lambda_n(t) ( \langle {\dot a_n(t)} |{\dot a_n(t)} \rangle +
 \langle {\dot b_n(t)} |{\dot b_n(t)} \rangle ) dt^2 \nonumber\\
&-& \big[ \sum_n \lambda_n(t) ( i \langle  a_n(t) |{\dot a_n(t)} \rangle + 
i \langle  b_n(t) |{\dot b_n(t)} \rangle) \big]^2 ~dt^2  \nonumber\\
&-& 2 \sum_{nm} \sqrt{\lambda_n(t)\lambda_m(t)} \langle  a_m(t) |{\dot a_n(t)} 
\rangle \langle  b_m(t) |{\dot b_n(t)} \rangle
+ \sum_n \frac{d \lambda_n(t)^2} {4 \lambda_n(t)} .
\end{eqnarray}
This metric is $U(1)$ gauge invariant and is independent of the 
detailed dynamics of the systems. This is so, because, there can be 
many non-local Hamiltonians which may give the same path in ${\cal P}$.
If the state is not entangled (one of the Schmidt number 
is one and others are zero), then $|\Psi(t) \rangle$ may be written as 
$|\Psi(t) \rangle = |a (t)\rangle |b (t) \rangle$ and 
the metric is then given by $dS^2 = dS_1^2 + dS_2^2 $, 
where $dS_i^2, (i=1,2)$ are the Fubini-Study metrics for the 
individual subsystems. In the above expression, the second and third terms 
represent the effect of entanglement on the metric structure.

Let us apply the nonlocal Hamiltonian for a time period $T$ and 
consider the quantum evolution 
$|\Psi(0) \rangle \rightarrow |\Psi(T) \rangle$. Suppose that the 
state $|\Psi(T) \rangle$ is our target state.
The total distance travelled by the quantum system in going from the initial 
state to the target state (as measured by the Fubini-Study 
metric) is given by 
$S = 2 \int_{\rm C} \frac{\Delta H}{\hbar}~~ dt = 2 Vt$, where $V$ is given 
in (3). This is typically a longer 
path in the projective Hilbert space of the entangled system.
Now, consider the shortest path $S_0$ 
between the initial and the target state. It is given by
\begin{eqnarray}
|\langle \Psi(0)|\Psi(T) \rangle |^2 = \cos^2 \frac{S_0}{2}.
\end{eqnarray}
Since the actual distance $S$ is greater than or equal to the shortest distance
$S_0$ (the Anandan-Aharonov version of the uncertainty relation) we have

\begin{eqnarray}
\Delta H~~ T \ge \frac{\hbar S_0}{2}.
\end{eqnarray}
From the above geometric relation we have a bound on the time required to reach
a target entangled state which is given by
\begin{eqnarray}
T \ge \frac{\hbar}{\Delta H} \cos^{-1} |\langle \Psi(0)|\Psi(T) \rangle |.
\end{eqnarray}

Therefore, to minimize the waiting time, we have to evolve the system 
faster and minimise the shortest path.
However, if the initial and target states are fixed, then the waiting time 
can be reduced by evolving the system faster. It may be noted that the above 
equation is valid both for qubits and qudits as well as for multiparticle 
quantum systems.

\section{Optimal time for two-qubit state}

In this section, we discuss the optimal time required to reach a 
two-qubit maximally entangled state starting from a pure product state and 
some partial entangled state.
Further, we will illustrate how the presence of initial entanglement helps us
to minimize the waiting time.

Let the initial state $|\Psi(0) \rangle$ of two qubits be 
\begin{eqnarray}
|\Psi_E\rangle = \sqrt p |0 \rangle | 1\rangle + \sqrt{1-p} 
| 1 \rangle | 0 \rangle , 
\end{eqnarray}
where $0 \le p \le 1/2$.
A general two qubit Hamiltonian can be written as

\begin{eqnarray}
H = \sum_{i=1}^3 \alpha_i \sigma_i \otimes I + \sum_{j=1}^3 
\beta_j I \otimes \sigma_j
+ \sum_{i,j=1}^3 \gamma_{ij} \sigma_i \otimes \sigma_j, 
\end{eqnarray}
where $\alpha_i, \beta_i$ are real numbers and $\gamma$ is a real matrix 
and $\sigma$'s are the usual Pauli matrices.
Supplementing the evolution operator with local unitary operators, we can 
rewrite the Hamiltonian as
$H  = \sum_{k=1}^3 \mu_k \sigma_k \otimes \sigma_k$.
where $\mu_1 \ge \mu_2 \ge \mu_3 \ge 0 $ are the sorted singular values of the 
matrix $\gamma$ \cite{dur}.
Let $|\Psi(0) \rangle$  evolves under a general two qubit Hamiltonian.
So the quantum evolution can be written as

\begin{eqnarray}
|\Psi_E \rangle \rightarrow |\Psi_T \rangle = 
e^{-i HT/\hbar}  |\Psi_E \rangle =  
e^{- i T \sum_{k=1}^3 \mu_k \sigma_k \otimes \sigma_k}|\Psi_E \rangle =
 |\Psi^{+}\rangle, 
\end{eqnarray}
where $|\Psi^{+}\rangle = \frac{1}{\sqrt 2} (|0 \rangle |1 \rangle + 
|1 \rangle|0 \rangle) $ 
is a maximally entangled state.
The speed with which two qubit state evolves in time is given by 
\begin{eqnarray}
V = \frac{ 2 \Delta H}{\hbar} = (\mu_1 + \mu_2) \sqrt{1-4p(1-p)}.
\end{eqnarray}

When the target is a maximally entangled state, 
then the bound on time is given by
\begin{eqnarray}
T \ge \frac{ S_0}{2 (\mu_1 + \mu_2) \sqrt{1-4p(1-p)} }, 
\end{eqnarray}
where $S_0 = 2 \cos^{-1}(\sqrt{p} + \sqrt{1-p})/\sqrt{2}$.
For the choice of best initial entangled state, we have 
$p =p_0 \approx 0.0832$ \cite{dur}.
This state has entanglement $E(\Psi_E) \approx 0.413$ ebit.
To increase the entanglement from $0.413$ ebit to one ebit, we must wait for a 
minimum time given by  $T_{\rm min} = \frac{0.5911}{(\mu_1 + \mu_2)}$.

Next, we ask: does the presence of initial entanglement help to reduce the 
waiting time? The answer is indeed yes.
Instead of an initial entangled state, if we start from a product state 
$|\Psi(0) \rangle = |0 \rangle | 1 \rangle$ and apply the unitary operator
$U(T)$ to reach a maximally entangled state (which is our target state), then 
the bound is given by 
$T \ge \frac{S_0}{(\mu_1 + \mu_2)}$,
where $S_0 = 2 \cos^{-1} (|\langle 0 1|\Psi^+\rangle |)$
The minimum time is given by $T_{\rm min} = \frac{0.7854}{(\mu_1 + \mu_2)}$.
This shows that if we start from a product state and want to reach a 
maximally entangled state, then we will have to wait longer. 
This also illustrates 
the role of initial condition (or initial entanglement) in the 
brachistocrone problem.


\section{ Bound on Time average of Entanglement Rate}

The study of dynamics of entanglement in quantum systems is an ongoing area of
research. Given a nonlocal 
Hamiltonian, it is important to know how well one can create entangled 
states. In Ref. \cite{dur}, 
it was found that it is better to start with initial entangled state and 
best initial entanglement is independent of the physical process. In this 
section, we will show that, in a bilocal rotating frame, the entangling rate 
is in fact a geometric quantity.

To quantify the entanglement production, one can define the entanglement rate 
$\Gamma(t) = \frac{dE(\Psi(t))}{dt}$, where $E(\Psi)$ is some 
entanglement measure for the state $|\Psi(t)\rangle $.
If we consider von Neumann entropy $E = -{\rm tr}(\rho(t) \log \rho(t))$ 
as our entanglement measure, then we have

\begin{eqnarray}
\Gamma(t) &=& - {\rm tr}({\dot \rho(t)} \log {\rho(t)}) 
=  -\sum_n \frac{d \lambda_n(t)}{dt} \log {\lambda_n(t)} \nonumber\\
&=& - 2\sum_{n,m=1}^N  \sqrt{\lambda_n(t) \lambda_m(t)}~~\log \lambda_n(t)  
h_{nm}(t),
\end{eqnarray}
where $h_{nm}(t) = 
{\rm Im}[\langle a_n(t)|\langle b_n(t)|H|a_m(t) \rangle |b_m(t) \rangle ]$ 
is called entangling capability of the nonlocal Hamiltonian.
In recent times, there have been various attempts to give bound on 
the entanglement rate. For example, 
upper bound on $\Gamma$ in the presence of local ancillas was proposed by 
Childs {\it et al} \cite{child}.
Wang and Sanders have shown that $\Gamma \le \beta \approx 1.9123$ 
for any product Hamiltonian $H = X \otimes X$, where $X$ is a self-inverse 
Hamiltonian \cite{wang}. Bennett {\it et al} have shown that 
$\Gamma(H) \le c d^4 || H||$, where $d$ is number of Schmidt coefficients, and
$c =O(1)$ \cite{ben}. Childs {\it et al} have shown that any two product 
Hamiltonians
can simulate each other \cite{childs}. Bandyopadhyay and Lidar have 
investigated the 
effect of noise on the entangling capacities of two-qubit Hamiltonians 
\cite{bando}. Very recently, Lari {\it et al} have investigated the 
entanglement rate for two qubits, two qutrits and three qubits 
using a geometric measure of entanglement which holds for multiparticle 
systems \cite{lari}.   
However, in the literature there is no universal bound for entanglement rate. 
What we plan in this paper is to provide a universal bound on the 
{\em time average of entanglement rate}.

Before, giving a bound on the time average of entanglement rate, 
we will show that in some 
special cases, the entangling capability is directly related to the speed of 
the transportation of the entangled quantum system. Once we apply the 
non-local Hamiltonian, the state at a later time is given by Eq(1). Suppose, 
it is possible to perform time-dependent local unitary transformations 
and make the 
Schmidt basis $\{|a_n(t)\rangle \}$ and $\{|b_n(t)\rangle \}$ time independent. 
This can be imagined as the state of the entangled system in a bilocal 
rotating frame. Since local unitaries cannot change the entanglement, we can
define a state 

\begin{eqnarray}
|\Psi_R(t) \rangle = U^{\dagger}(t) \otimes  V^{\dagger}(t) |\Psi(t) \rangle
= \sum_{n=1}^N \sqrt{\lambda_n(t)} |a_n \rangle |b_n \rangle, 
\end{eqnarray}
where $|a_n(t) \rangle =  U(t) |a_n \rangle$ and  $|b_n(t) \rangle= 
V(t) |b_n \rangle$. Thus, at any given instant of time the entanglement content 
of $|\Psi_R(t) \rangle$ and $|\Psi(t) \rangle$ are the same, i.e., 
$E(\Psi_R(t)) = E(\Psi(t))$

Now, let us look at the (squared) speed at which the state  
$|\Psi_R(t) \rangle$ evolves in time. It is given by
\begin{eqnarray}
v^2 = 4 \frac{\Delta H^2}{\hbar^2} = \sum_n \frac{1}{\lambda_n(t)}
(\frac{d \lambda_n(t)}{dt})^2.
\end{eqnarray}
For two-qubits with the general non-local Hamiltonian one has 
\begin{eqnarray}
\frac{dp}{dt} = 2 \sqrt{p(1-p)} 
{\rm Im}[\langle a_0|\langle b_0|H|a_1  \rangle |b_1 \rangle],
\end{eqnarray}
where $\lambda_1 =p$ and $\lambda_2 =(1-p)$.
Therefore, the speed of the system can be expressed as
\begin{eqnarray}
v^2 =  \frac{{\dot p(t)}^2}{p(t)(1-p(t))} = 
4 {\rm Im}[\langle a_0|\langle b_0|H|a_1  \rangle |b_1 \rangle ]^2.
\end{eqnarray}
Note that $h = {\rm Im}[\langle a_0|\langle b_0|H|a_1  \rangle |b_1 \rangle$ 
and $h_{\rm max}$ is entangling capability of the non-local Hamiltonian. 
Therefore, from (17) we have $V = 2 h$. Thus, $V_{\rm max}$, 
the maximum speed of the entangled 
system during quantum evolution is directly related to 
the entangling capability of two-qubit Hamiltonian. This shows that 
in a bilocal rotating frame the entangling capability is a geometric quantity.

Next, we will show that one can give a universal bound on the 
{\em time average of entanglement} for any initial state and 
general nonlocal  Hamiltonians. Consider the evolution of a 
bipartite system $|\Psi(0) \rangle \rightarrow |\Psi(T) \rangle$ during an 
interval $[0, T]$. We define the time average of entanglement 
${\bar \Gamma}$ as 

\begin{equation}
{\bar \Gamma} = \frac{1}{T} \int_0^T  \Gamma(t) dt 
= -  \frac{1}{T} \sum_n \int_0^T \log {\lambda_n(t)}~ d \lambda_n(t).
\end{equation}

Note that during evolution of a composite system the change in 
the entanglement content can never
exceed $\log N$, i.e., $ \delta E = E(T) - E(0) \le \log N$, where $N$ is 
number of Schmidt coefficients in the entangled state. This implies that
${\bar \Gamma} \le \frac{\log N}{T}$. Now, using the geometric uncertainty 
relation for time and energy fluctuation, we find 
\begin{eqnarray}
{\bar \Gamma} \le 2\log N \frac{\Delta H}{\hbar S_0}.
\end{eqnarray}
Thus, the maximal value of time average of entanglement rate depends on 
the maximum Schmidt rank, on the quantum fluctuation in the nonlocal
Hamiltonian and inversely on the shortest geodesic path. For a given 
system, the time average of entanglement does not depend on the 
particular Hamiltonian, rather on the fluctuation in the Hamiltonian. 
There can be infinite number of nonlocal Hamiltonians, yet they all may 
give same average entanglement rate. For two qubits, starting with the initial 
state as given in (8), the 
maximal time average of entanglement is given by
\begin{eqnarray}
{\bar \Gamma}_{\rm max} =  \frac{(\mu_1 + \mu_2)}{S_0} \sqrt{1- 4p(1-p)}.
\end{eqnarray}

It may be mentioned that it is not yet known whether the maximal 
entanglement and disentanglement rates of a nonlocal Hamiltonian are 
the same, i.e. if $\Gamma(H) =
\Gamma(-H)$? We would like to point out that even though we do not know 
complete answer to the above question, our result suggests that 
the maximum attainable value for the time averaged entanglement rate and 
time averaged disentanglement rates are identical. The maximum of  
time average entanglement rate is given by 
${\bar \Gamma}_{max}(H) =  2\log N \frac{\Delta H}{\hbar S_0}$. 
Since it depends on the quantum mechanical uncertainty, we have 
$\Delta H = \Delta (-H)$ and because of the law of reciprocity of 
quantum mechanical probability ($S_0$ is same during entangling and 
disentangling evolutions) we have ${\bar \Gamma}_{max}(H) =
{\bar \Gamma}_{max}(-H)$. This is another non-trivial consequence of 
our geometric approach to the problem.

\section{Minimum time for composite Hamiltonian}

Now, we ask what happens to the minimal time if we consider a time evolution 
under one Hamiltonian instead of another. In particular, if the later is 
build up from a composition of Hamiltonians then how does the minimum time 
behave. The answer to this question may be useful in quantum computing 
algorithms, where one tries to design sequence of unitary operators via 
the application of suitable Hamiltonians. It is known that if we have 
the ability to perform evolution under a Hamiltonian $H$ (say) and have the
ability to apply unitaries $U_k$, it is possible to simulate evolution 
according to a 
composite Hamiltonian like $H'=  \sum_k \alpha_k U_k H U_k^{\dagger}$,
where $\alpha_k$'s are real numbers \cite{niel}.
Under such an evolution we will have
\begin{eqnarray}
T \ge \frac{\hbar}{2 \Delta H'} \cos^{-1} 
| \langle \Psi(0)|U(T)'|\Psi(0) \rangle |,
\end{eqnarray}
where $U(T) = \exp(- i H' T/\hbar)$ and $\Delta H'$ is the 
quantum fluctuation in the composite Hamiltonian. Suppose, that the initial and
the final states are fixed, i.e., we start with same initial state and 
want to reach a desired target state with the new Hamiltonian $H'$.  
Then, the shortest path $S_0$ is again the same. Now, using the 
convexity property of quantum mechanical 
uncertainty \cite{akp2}, we have 
\begin{eqnarray}
\Delta H' = \Delta(\sum_k  \alpha_k U_k H U_k^{\dagger}) \le 
\sum_k \alpha_k \Delta (U_k H U_k^{\dagger}).
\end{eqnarray}
This tells us that mixing different Hamiltonians decreases the 
quantum mechanical uncertainty. If we define
$H_k=  U_k H U_k^{\dagger}$, then the minimum time for the new Hamiltonian 
satisfies
\begin{eqnarray}
T \ge \frac{\hbar S_0}{2 \Delta H'}  \ge \frac{\hbar S_0}
{2 \sum_k \alpha_k \Delta H_k }.
\end{eqnarray}

This also gives a composition law for the minimum time for 
entangled brachistocrone
problem under composite Hamiltonian.
To illustrate this, consider a composite Hamiltonian $H'$ that consists of two 
Hamiltonians $H_1$ and $H_2$. If we evolve the system only under $H_1$ then the
minimum time to reach the target entangled state will be 
$T_{1_{min}} = \frac{\hbar S_0}{2\Delta H_1}$. Similarly, 
if we evolve the system only under $H_2$ then the
minimum time to reach the target entangled state will be 
$T_{2_{min}} = \frac{\hbar S_0}{2\Delta H_2}$. Now, 
if we evolve the system under $H'= \alpha_1 H_1 + \alpha_2 H_2$ then the
minimum time to reach the target entangled state will be 
\begin{eqnarray}
T_{min} =  \frac{T_{min1} T_{min2}}
{ \alpha_1 T_{min2} + \alpha_2  T_{min1}}.
\end{eqnarray}

To illustrate this composition law consider the initial state as given in (8).
Now suppose, we apply the Hamiltonian $H_1 = \mu_1 \sigma_1 \otimes \sigma_1$ 
and evolve the system for sometime to reach the target entangled state. 
In this case, the minimum time we need to wait is given by  
$T_{1_{min}} = \frac{\hbar S_0}{2 \mu_1 \sqrt{1-4p(1-p)} }$. If we apply instead 
$H_2 = \mu_2 \sigma_2 \otimes \sigma_2$ 
and evolve the system to reach the target entangled state, then 
the minimum time we need to wait is given by  
$T_{2_{min}} = \frac{\hbar S_0}{2 \mu_2 \sqrt{1-4p(1-p)} }$. Now, we apply 
$H = \alpha_1 H_1 + \alpha_2 H_2$ 
and evolve the system for sometime to reach the same target entangled state. 
In this case, the minimum time we need to wait is given by  
$T_{min} = \frac{\hbar S_0}{2 (\alpha_1 \mu_1 +\alpha_2 \mu_2) 
\sqrt{1-4p(1-p)} }$. This later quantity indeed satisfies the composition law
as given in (24). In fact, with a suitable controlled experiment, one 
can even test the composition law for the minimum time.
 
\section{Conclusion}

We have investigated the brachistochrone problem for entangled states. 
This answers the question that we have raised in the beginning, namely, 
given an arbitrary product state and 
general non-local Hamiltonian, what is the minimum time for creating a desired
entangled state? We have shown that the minimum time depends inversely on the 
speed of the evolution of the quantum system. This is independent of the 
particular entanglement measure and holds even for multiparticle entangled
systems. We have shown that for two qubits 
the presence of initial entanglement does help in minimising the waiting time.
In particular, we find that if we are given an option of starting from 
a product state and an entangled state, and wants to
reach the maximally entangled state, then it is better to start from some 
already initial entangled state. Because, in the later case the time required 
is less. We have shown that the time average of entanglement rate depends 
on the Schmidt rank and on the fluctuation in the non-local Hamiltonian. 
We have also shown 
that in a bilocal rotating frame the entangling capability is also 
directly related to the 
speed of the evolution of the entangled states. This shows that the 
entangling capability of a non-local Hamiltonian is actually a geometric 
quantity. Furthermore, we have proved 
a composition law for the minimum time when the system evolves under composite 
Hamiltonians. We hope that these findings will be of interest in 
understanding the entanglement rate and entangling capabilities of 
non-local Hamiltonians and in real experimental situations.
In future, it will be interesting to see if all these findings 
also hold for qudits and multiparticle systems.


\end{document}